\input harvmac
\def\O{{\cal O}}

\def\nm{{\nu_\mu}}
\def\ne{{\nu_e}}
\def\gsim{{~\raise.15em\hbox{$>$}\kern-.85em
          \lower.35em\hbox{$\sim$}~}}
\def\lsim{{~\raise.15em\hbox{$<$}\kern-.85em
          \lower.35em\hbox{$\sim$}~}}
\def\lpp{\lambda^{\prime\prime}} 

\def\YGTitle#1#2{\nopagenumbers\abstractfont\hsize=\hstitle\rightline{#1}%
\vskip .4in\centerline{\titlefont #2}\abstractfont\vskip .3in\pageno=0}
\YGTitle{WIS-99/17/Apr-DPP, hep-ph/9904473}
{\vbox{\centerline{Lepton Parity in Supersymmetric Flavor Models}}}
\bigskip
\centerline{Galit Eyal and Yosef Nir}
\smallskip
\centerline{\it Department of Particle Physics,
 Weizmann Institute of Science, Rehovot 76100, Israel}
\bigskip
\bigskip
\baselineskip 18pt
\noindent
We investigate supersymmetric models where neither $R$ parity nor lepton 
number nor baryon number is imposed. The full high energy theory
has an exact horizontal $U(1)$ symmetry that is spontaneously broken. 
Quarks and Higgs fields carry integer horizontal charges but leptons 
carry half integer charges. Consequently, the effective low energy theory
has two special features: a $U(1)$ symmetry that is explicitly broken
by a small parameter, leading to selection rules, and an exact residual
$Z_2$ symmetry, that is lepton parity. As concerns neutrino parameters,
the $Z_2$ symmetry forbids contributions from $R_p$-violating couplings
and the $U(1)$ symmetry induces the required hierarchy. As concerns
baryon number violation, the $Z_2$ symmetry forbids proton decay
and the $U(1)$ symmetry provides sufficient suppression of
double nucleon decay and of neutron $-$ antineutron oscillations.
\bigskip
 
\baselineskip 18pt
\leftskip=0cm\rightskip=0cm
 
\Date{}

\newsec{Introduction}
In contrast to the Standard Model (SM), the Supersymmetric Standard Model 
(SSM) does not have accidental lepton- ($L$) and baryon-number ($B$) 
symmetries.
This situation leads to severe phenomenological problems, 
{\it e.g.} fast proton decay and large neutrino masses.
The standard solution to this problem is to impose a discrete symmetry, $R_p$. 
The SSM with $R_p$ does have $L$ and $B$ as accidental symmetries. 

Both the SM and the SSM provide no explanation for the smallness and 
hierarchy in the Yukawa couplings. One way of explaining the flavor puzzle 
is to impose an approximate horizontal symmetry. Such symmetries suppress 
not only the Yukawa couplings but also the $B$ and $L$  violating terms of 
the SSM
\nref\HiKa{I. Hinchliffe and T. Kaeding, Phys. Rev. D47 (1993) 279.}%
\nref\BHNi{V. Ben-Hamo and Y. Nir,
  Phys. Lett. B339 (1994) 77, hep-ph/9408315.}%
\nref\BGNN{T. Banks, Y. Grossman, E. Nardi and Y. Nir,
  Phys. Rev. D52 (1995) 5319, hep-ph/9505248.}%
\nref\SmVi{A.Y. Smirnov and F. Vissani,
 Nucl. Phys. B460 (1996) 37, hep-ph/9506416.}%
\nref\CHM{C.D. Carone, L.J. Hall and H. Murayama, 
 Phys. Rev. D53 (1996) 6282, hep-ph/9512399;
 Phys. Rev. D54 (1996) 2328, hep-ph/9602364.}%
\nref\BLR{P. Binetruy, S. Lavignac and P. Ramond,
 Nucl. Phys. B477 (1996) 353, hep-ph/9601243.}%
\nref\ChLu{E.J. Chun and A. Lukas,
 Phys. Lett. B387 (1996) 99, hep-ph/9605377.}%
\nref\BoGNN{F.M. Borzumati, Y. Grossman, E. Nardi and Y. Nir,
 Phys. Lett. B384 (1996) 123, hep-ph/9606251.}%
\nref\ENar{E. Nardi, Phys. Rev. D55 (1997) 5772, hep-ph/9610540.}%
\nref\CCKi{K. Choi, E.J. Chun and H. Kim,
 Phys. Lett. B394 (1997) 89, hep-ph/9611293.}%
\nref\BILR{P. Binetruy, N. Irges, S. Lavignac and P. Ramond,
 Phys. Lett. B403 (1997) 38, hep-ph/9612442.}%
\nref\Bhat{G. Bhattacharyya, Phys. Rev. D57 (1998) 3944, hep-ph/9707297.}%
\nref\BDLS{P. Binetruy, E. Dudas, S. Lavignac and C.A. Savoy,
 Phys. Lett. B422 (1998) 171, hep-ph/9711517.}%
\nref\ILR{N. Irges, S. Lavignac and P. Ramond,
 Phys. Rev. D58 (1998) 035003, hep-ph/9802334.}%
\nref\ELR{J. Ellis, S. Lola and G.G. Ross,
 Nucl. Phys. B526 (1998) 115, hep-ph/9803308.}%
\nref\CCH{K. Choi, E.J. Chun and K. Hwang, hep-ph/9811363.}%
\refs{\HiKa-\CCH}. Consequently, it is possible to construct viable 
supersymmetric models 
with horizontal symmetries and without $R_p$. The phenomenology of these models 
is very different from that of $R_p$-conserving models. In particular, the LSP 
is unstable, various $L$ and $B$ violating processes may occur at an observable
level, and an interesting pattern of neutrino masses is predicted.

It is not simple, however, to solve all problems of $L$ and $B$ violation by 
means of horizontal symmetries:
\item{(a)} Constraints from proton decay require uncomfortably large horizontal charges for quarks to sufficiently suppress the $B$ violating terms \BHNi;
\item{(b)} In models where the  $\mu$ terms are not aligned with the $B$ terms, 
constraints from neutrino masses require uncomfortably large horizontal charges 
for leptons to sufficiently suppress the $L$ violating terms 
\refs{\BGNN,\BoGNN}.

Therefore, most models with horizontal symmetries and without $R_p$ still 
impose baryon number symmetry and assume $\mu-B$ alignment at some high energy 
scale. In this work we show that it is not necessary to make these assumptions: 
one can construct viable supersymmetric models without $R$ parity, without 
lepton number, without baryon number and with horizontal charges that are not 
very large. The crucial point is that the horizontal $U(1)_H$ symmetry is not 
completely broken: a residual discrete symmetry, lepton parity, forbids proton
decay and aligns the $\mu$ and $B$ terms. The constraints on the baryon number
violating terms from double nucleon decay and neutron-antineutron oscillations 
are easily satisfied and interesting neutrino parameters can be accommodated 
naturally in these models.

The idea that lepton parity could arise from the spontaneous breaking of a 
horizontal symmetry was first suggested, to the best of our knowledge, in
\ref\IbRo{L.E. Ibanez and G.G. Ross, Nucl. Phys. B368 (1992) 3.}.
Explicit models, with a horizontal $U(1)$ symmetry and a residual
$Z_2$ (different from lepton parity), were presented in refs.
\refs{\CCKi,\CCH}.

The plan of this paper is as follows. In section 2, we define our notations.
In section 3 we present an explicit model and its predictions for
the Yukawa parameters in the quark and in the lepton sectors. We emphasize
that it is not easy to accommodate the parameters of the
MSW solution to the solar neutrino problem with a small mixing angle.
In section 4 we investigate the consequences of the residual
lepton parity on $R$-parity violating couplings. A summary is given
and various comments are made in section 5.

\newsec{Notations}
The matter supermultiplets are denoted in the following way:
\eqn\matrep{\eqalign{
&Q_i(3,2)_{+1/6},\ \ \ \bar u_i(\bar3,1)_{-2/3},\ \ \ 
\bar d_i(\bar3,1)_{+1/3},\cr 
&L_i(1,2)_{-1/2},\ \ \ \bar\ell_i(1,1)_{+1},\ \ \ N_i(1,1)_0,\cr
&\phi_u(1,2)_{+1/2},\ \ \ \phi_d(1,2)_{-1/2}.\cr}}
The $N_i$ supermultiplets are Standard Model singlets. Their masses are 
assumed to be much heavier than the electroweak breaking scale but lighter 
than the scale of $U(1)_H$ breaking. We denote this intermediate mass scale 
by $M$. Lepton number is violated by bilinear terms in the superpotential,
\eqn\muLV{\mu_i L_i\phi_u,}
and by trilinear terms in the superpotential,
\eqn\lambdaLV{\lambda_{ijk}L_iL_j\bar\ell_k+\lambda^\prime_{ijk}L_iQ_j\bar d_k.}
Baryon number is violated by trilinear terms in the superpotential,
\eqn\lambdaBV{\lambda^{\prime\prime}_{ijk}\bar u_i\bar d_j\bar d_k.}
There are also $L$ breaking supersymmetry breaking bilinear terms in the 
scalar potential:
\eqn\BLV{B_i L_i\phi_u,}
and
\eqn\msqLV{\tilde m^2_{i0}L_i^\dagger\phi_d,}
where here $L_i$, $\phi_d$ and $\phi_u$ denote scalar fields.

\newsec{The Yukawa Hierarchy}
\subsec{A Simple Model}
Consider a model with a horizontal symmetry $U(1)_H$. The symmetry is broken 
by two small parameters, $\lambda$ and $\bar\lambda$, to which we attribute 
$H$-charges of $+1$ and $-1$, respectively.  We give them equal values 
(so that the corresponding $D$ terms do not lead to supersymmetry breaking 
at a high scale). For concreteness we take $\lambda=\bar\lambda=0.2$. 
At low energies, we have then the following selection rules:
\item{a.} Terms in the superpotential and in the Kahler potential that 
carry an integer $H$-charge $n$ are suppressed by $\O(\lambda^{|n|})$.
\item{b.} Terms in the superpotential and in the Kahler potential that carry a 
non-integer charge vanish.

We set the $H$ charges of the matter fields as follows:
\eqn\matterHH{\matrix{\phi_u&\phi_d\cr(0)&(0)\cr}}
\eqn\matterqH{\matrix{Q_1&Q_2&Q_3&&\bar u_1&\bar u_2&\bar u_3&&\bar 
d_1&\bar d_2&\bar d_3\cr (3)&(2)&(0)&&(4)&(2)&(0)&&(4)&(3)&(3)\cr}}
\eqn\matterlH{\matrix{L_1&L_2&L_3&&\bar\ell_1&\bar\ell_2&\bar\ell_3&&
N_1&N_2&N_3\cr (7/2)&(-1/2)&(-1/2)&&(11/2)&(11/2)&(7/2)&&(-1/2)&(1/2)&(1/2)
\cr}}

The selection rules dictate then the following form for the
quark mass matrices:
\eqn\qmasmat{M_u\sim\vev{\phi_u}\pmatrix{\lambda^7&\lambda^5&\lambda^3\cr 
\lambda^6&\lambda^4&\lambda^2\cr \lambda^4&\lambda^2&1\cr},\ \ \ 
M_d\sim\vev{\phi_d}\pmatrix{\lambda^7&\lambda^6&\lambda^6\cr
\lambda^6&\lambda^5&\lambda^5\cr \lambda^4&\lambda^3&\lambda^3\cr}.}
In eq. \qmasmat\ and below, unknown coefficients of $\O(1)$
are not explicitly written. These mass matrices give order of magnitude 
estimates for the physical parameters (masses and mixing angles) that
are consistent with the experimental data (extrapolated to a high
energy scale):
\eqn\qphys{\eqalign{
&m_t/\vev{\phi_u}\sim1,\ \ \ m_c/m_t\sim\lambda^4,\ \ \ 
m_u/m_c\sim\lambda^3,\cr
&m_b/m_t\sim\lambda^3,\ \ \ m_s/m_b\sim\lambda^2,\ \ \ 
m_d/m_s\sim\lambda^2,\cr
&|V_{us}|\sim\lambda,\ \ \ |V_{cb}|\sim\lambda^2,\ \ \ 
|V_{ub}|\sim\lambda^3.\cr}}

For the charged leptons mass matrix $M_\ell$, the neutrino Dirac mass
matrix $M_\nu^{\rm Dirac}$, and the Majorana mass matrix for the
singlet neutrinos $M_N$, we have
\eqn\chargedl{M_\ell\sim\vev{\phi_d}\pmatrix{
\lambda^9&\lambda^9&\lambda^7\cr \lambda^5&\lambda^5&\lambda^3\cr
\lambda^5&\lambda^5&\lambda^3\cr},}
\eqn\neutrall{M_\nu^{\rm Dirac}\sim\vev{\phi_u}\pmatrix{
\lambda^3&\lambda^4&\lambda^4\cr \lambda&1&1\cr
\lambda&1&1\cr},\ \ \ M_N\sim M
\pmatrix{\lambda&1&1\cr 1&\lambda&\lambda\cr1&\lambda&\lambda\cr}.}
These matrices give the following order of magnitude estimates:
\eqn\lphys{\eqalign{
&m_\tau/\vev{\phi_d}\sim\lambda^3,\ \ \ m_\mu/m_\tau\sim\lambda^2,\ \ \ 
m_e/m_\mu\sim\lambda^4,\cr
&m_{\nu_3}/(\vev{\phi_u}^2/M)\sim1/\lambda,\ \ \ 
m_{\nu_2}/m_{\nu_3}\sim\lambda^2,\ \ \ 
m_{\nu_1}/m_{\nu_2}\sim\lambda^4,\cr
&|V_{e\nu_2}|\sim\lambda^2,\ \ \ |V_{\mu\nu_3}|\sim1,\ \ \ 
|V_{e\nu_3}|\sim\lambda^4.\cr}}
The neutrino parameters fit the atmospheric neutrino data and the
small mixing angle (SMA) MSW solution of the solar neutrino problem
\ref\BHSSW{R. Barbieri, L.J. Hall, D. Smith, A. Strumia and N. Weiner,
 JHEP 12 (1998) 017, hep-ph/9807235.}.

\subsec{The Neutrino Mass Hierarchy}
As concerns neutrino parameters, the most predictive class of models is the 
one where $s_{23}$ and $\Delta m^2_{\rm SN}/\Delta m^2_{\rm AN}$ depend
only on the horizontal charges of $L_2$, $L_3$, $\bar\ell_2$ and $\bar\ell_3$
\nref\GrNi{Y. Grossman and Y. Nir,
 Nucl. Phys. B448 (1995) 30, hep-ph/9502418.}%
\nref\GNS{Y. Grossman, Y. Nir and Y. Shadmi,
 JHEP 10 (1998) 007, hep-ph/9808355.}%
\refs{\GrNi,\GNS}. We call such models, where the horizontal charges of 
neither the first generation nor sterile neutrinos affect the above 
parameters, (2,0) models. Models with $n_a$ active and $n_s$ sterile 
neutrinos are denoted by ($n_a,n_s$). It was proven in
\ref\NiSh{Y. Nir and Y. Shadmi, hep-ph/9902293.}\
that in ($2,n_s\leq2$) models, for neutrinos with large mixing,
$s_{23}\sim1$, we have $m_2/m_3\sim\lambda^{4n}$ 
($\Delta m^2_{\rm SN}/\Delta m^2_{\rm AN}\sim\lambda^{8n}$). Therefore,
the MSW solutions, which require $\Delta m^2_{\rm SN}/\Delta m^2_{\rm AN}
\sim\lambda^2-\lambda^4$, cannot be accommodated in this framework. The LMA 
solution can be achieved in $n_a=3$ models (for any $n_s$) but the SMA 
solution requires $n_s\geq3$. (For an $n_s=3$ model, see, for example,
\ref\AlFe{G. Altarelli and F. Feruglio, JHEP 11 (1998) 021, hep-ph/9809596;
 Phys. Lett. B451 (1999) 388, hep-ph/9812475.}.) 
This means a loss of predictive power, particularly in comparison with
$n_s=0$ models.

The proof in ref. \NiSh\ referred to models with only integer horizontal
charges (in units of the charge of the breaking parameters). The question
arises then whether one can have a hierarchy for
$\Delta m^2_{\rm SN}/\Delta m^2_{\rm AN}$ that is milder than $\lambda^{8n}$
in models where leptons carry half-integer charge even for $n_s\leq2$.
We now show that the answer to this question is negative.

Consider (2,0) models with $H(L_2)\neq H(L_3)$. The large mixing can be
obtained from the charged lepton mass matrix if the following
condition is fulfilled \NiSh:
\eqn\larmix{H(L_2)+H(L_3)=-2H(\bar\ell_3).}
The hierarchy is given by
\eqn\twzehie{{m(\nu_2)\over m(\nu_3)}\sim\lambda^{2|H(L_2)+H(L_3)|-4|H(L_3)|}.}
From \larmix\ and \twzehie\ we find
\eqn\ATSNtz{{\Delta m^2_{\rm SN}\over\Delta m^2_{\rm AN}}\sim
\lambda^{8(|H(\bar\ell_3)|-|H(L_3)|)}.}
Since $H(\bar\ell_3)$ and $H(L_3)$ are both half-integers, the
difference $|H(\bar\ell_3)|-|H(L_3)|$ is an integer and the hierarchy
is $\lambda^{8n}$.

The same statement
($\Delta m^2_{\rm SN}/\Delta m^2_{\rm AN}\sim\lambda^{8n}$) 
holds also in (2,2) models. (The proof for that is quite lengthy;
it follows lines similar to Appendix A in \NiSh\ and we do not
present it here.) We conclude then that models
where leptons carry half-integer charges do not provide
new ways to achieve a mild hierarchy between
$\Delta m^2_{\rm SN}$ and $\Delta m^2_{\rm AN}$. For the MSW
solutions we have either the LMA solution with $\ne-\nm$ forming
a pseudo-Dirac neutrino or at least three sterile neutrinos
playing a role in the light neutrino flavor parameters.  
 
\newsec{$L$ and $B$ Violation}
The model described above has an exact $Z_2$ symmetry, that is lepton parity.
This symmetry follows from the selection rules. But it can be understood
in a more intuitive way from the full high energy theory. We assume here
that our low energy effective theory given in the previous section comes from
a (supersymmetric version
\ref\LNS{M. Leurer, Y. Nir and N. Seiberg,
 Nucl. Phys. B398 (1993) 319, hep-ph/9212278.}\ 
of) the Froggatt-Nielsen mechanism
\ref\FrNi{C.D. Froggatt and H.B. Nielsen, Nucl. Phys. B147 (1979) 277.}.
The full high energy theory has an exact $U(1)_H$ symmetry that is 
spontaneously broken by the VEVs of two scalar fields, $\phi$ and
$\bar\phi$, of $H$-charges $+1$ and $-1$, respectively. Quarks and
leptons in vector representation of the SM gauge group and of $U(1)_H$,
and with very heavy masses $M_{\rm FN}$ communicate the information about
the breaking to the SSM fields ($\lambda=\vev{\phi}/M_{\rm FN}$ and
$\bar\lambda=\vev{\bar\phi}/M_{\rm FN}$).
 
The $U(1)_H$ symmetry has a $Z_2$ subgroup where all
fields that carry half-integer $H$-charges are odd, while all those
that carry integer $H$-charges are even. This symmetry is not broken
by $\vev{\phi}$ and $\vev{\bar\phi}$ since $\phi$ and $\bar\phi$ are
$Z_2$ even. Our choice of $H$-charges is such that all leptons
($L_i$, $\bar\ell_i$ and $N_i$) carry half-integer charges and therefore
are $Z_2$-odd. All other fields (quarks and Higgs fields) carry integer
charges and therefore are $Z_2$-even. We can identify the exact residual
symmetry then as lepton parity. 

Lepton parity is very powerful in relaxing the phenomenological problems
that arise in supersymmetric models without $R_p$. In particular, it forbids
the bilinear $\mu$ terms of eq. \muLV, the $B$ terms of eq. \BLV, the
$\tilde m^2$ terms of eq. \msqLV, and the trilinear terms of eq. \lambdaLV. 
The only allowed renormalizable $R_p$ violating terms are the baryon number 
violating couplings of eq. \lambdaBV.

This situation has two interesting consequences:

(i) Similarly to $R_p$ conserving models, the only allowed $\mu$ term is 
$\mu\phi_u\phi_d$ and the only allowed $B$ term is $B\phi_u\phi_d$.
The $\mu$ and $B$ terms are then aligned. Furthermore, the
mass-squared matrix for the scalar $(1,2)_{-1/2}$ fields can be separated
to two blocks, a $3\times3$ block for the three slepton fields and a single
term for $\phi_d$. Therefore there will be no renormalizable tree-level
contribution to neutrino masses
\nref\HaSu{L.J. Hall and M. Suzuki, Nucl. Phys. B231 (1984) 419.}%
\refs{\HaSu,\BGNN}. Consequently, the very large $H$ charges that are
needed to achieve precise $\mu$-$B$ alignment are not necessary here. 

Since the $\lambda_{ijk}$ and $\lambda^\prime_{ijk}$
couplings vanish, there will also be no $R_p$ breaking loop contributions
to neutrino masses. On the other hand, the usual see-saw contributions
which break lepton number by two units are allowed. This justifies why
we considered \neutrall\ as the only source for neutrino masses.  

(ii) Since processes that violate lepton number by one unit are forbidden,
the proton is stable. (We assume here that there is no fermion that is
lighter than the proton and does not carry lepton number.)
The most severe constraints on baryon number violating
couplings are then easily satisfied.

On the other hand, the $\lambda^{\prime\prime}$ couplings of eq. \lambdaBV\
contribute to double nucleon decay, to neutron-antineutron oscillations and to 
other rare processes
\nref\Zwir{F. Zwirner, Phys. Lett. B132 (1983) 103.}%
\nref\BaMa{R. Barbieri and A. Masiero, Nucl. Phys. B267 (1986) 679.}%
\nref\BGH{V. Barger, G.F. Giudice and T. Han, Phys. Rev. D40 (1989) 2987.}%
\nref\GoSh{J.L. Goity and M. Sher,
 Phys. Lett. B346 (1995) 69, hep-ph/9412208, Erratum-ibid. B385 (1996) 500.}%
\nref\BCS{G. Bhattacharyya, D. Choudhury and K. Sridhar,
 Phys. Lett. B355 (1995) 193, hep-ph/9504314.}%
\nref\CRS{C.E. Carlson, P. Roy and M. Sher,
 Phys. Lett. B357 (1995) 99, hep-ph/9506328.}%
\nref\ChKe{D. Chang and W.Y. Keung,
 Phys. Lett. B389 (1996) 294, hep-ph/9608313.}%
\refs{\Zwir-\ChKe}. 

The non-observation of baryon number violating processes gives strong 
constraints on all the $\lambda^{\prime\prime}$ couplings, {\it e.g.}
\eqn\neannea{\lpp_{112}\ \leq\ 10^{-6},\ \ \ 
\lpp_{113}\ \leq\ 5\times10^{-3}.}
The first bound comes from double nucleon decay and the second from
neutron-antineutron oscillations, and they correspond to a typical
supersymmetric mass $\tilde m\sim300\ GeV$. In our models, all the relevant 
constraints are satisfied since the $\lambda^{\prime\prime}$ couplings are 
suppressed by the selection rules related to the broken $U(1)$. Explicitly, 
our choice of $H$-charges in eq. \matterqH\ leads to the following
order of magnitude estimates:
\eqn\suppresspp{\eqalign{
\lpp_{112}&\sim\lpp_{113}\sim\lambda^{11},
\ \ \ \lpp_{123}\sim\lambda^{10},\cr
\lpp_{212}&\sim\lpp_{213}\sim\lambda^{9},
\ \ \ \lpp_{223}\sim\lambda^{8},\cr
\lpp_{312}&\sim\lpp_{313}\sim\lambda^{7},
\ \ \ \lpp_{323}\sim\lambda^{6}.\cr}}
The value that is closest to the bound is that of $\lpp_{112}$,
predicting double nucleon decay at a rate that, for $\tilde m\sim100\ GeV$,
is four orders of magnitude below the present bound. 

Note, however, that reasonable variations on our model can easily
give larger $\lpp$ couplings and allow the upper bound on double
nucleon decay to be saturated. For example, replacing the $H$ charges
in eq. \matterqH\ with a linear combination of $H$ and baryon number $B$
($H^\prime=a_1 H+a_2B$) does not affect the $B$ conserving quantities 
and, in particular, the mass matrices \qmasmat, \chargedl\ and \neutrall, 
but does affect (and, in particular, can enhance) the $\lpp$ couplings in 
\suppresspp. The couplings could also be affected by $\tan\beta$.
Our choice of charges corresponds to $\tan\beta\sim1$. 
But for large $\tan\beta$ and the same choice of $H$-charges for $\phi_d$
and $Q_i$, the $\lpp$ couplings are enhanced by $\tan^2\beta$. We conclude 
that, within our framework, baryon number violating processes could occur 
at observable rates.

\newsec{Summary and Comments}
In the framework of supersymmetric models, horizontal $U(1)$ symmetries
can lead to many interesting consequences, the most important being
a natural explanation of the smallness and hierarchy in the Yukawa parameters.
We have investigated a particular class of models, where the horizontal
$U(1)$ is the {\it only} symmetry imposed on the model beyond
supersymmetry and the Standard Model gauge symmetry. In particular, we
have imposed neither $R$-parity, nor lepton number nor baryon number.
Usually, such models can be made viable only at the price of assigning
uncomfortably large horizontal charges to various matter fields.
It is possible, however, that the horizontal symmetry leads to
exact lepton parity at low energy. The constraints that usually require
the large charges are irrelevant because proton decay is forbidden
and because mixing between neutrinos and neutralinos is forbidden.
The remaining constraints from double nucleon decay and from neutrino 
masses are easily satisfied by the selection rules of the broken $U(1)$.

Our emphasis here has been put on lepton and baryon number
violation. Therefore, we have ignored two other aspects of our framework. 
First, we did not insist that the horizontal symmetry solves the 
supersymmetric flavor problem. It is actually impossible to sufficiently
suppress the supersymmetric contributions to flavor changing neutral
currents by means of a single horizontal $U(1)$ symmetry. It is possible
that this problem is solved by a different mechanism. For example,
squarks and sleptons could be degenerate as a result of either
dilaton dominance in Supersymmetry breaking or a universal
gaugino contribution in the RGE (for a recent discussion, see
\ref\Eyal{G. Eyal, hep-ph/9903423.}).
Alternatively, one could complicate the model by employing a $U(1)\times
U(1)$ symmetry to achieve alignment
\nref\NiSe{Y. Nir and N. Seiberg,
 Phys. Lett. B309 (1993) 337, hep-ph/9304307.}%
\nref\LNSb{M. Leurer, Y. Nir and N. Seiberg,
 Nucl. Phys. B420 (1994) 468, hep-ph/9310320.}%
\refs{\NiSe,\LNSb}. In either case, the implications
for the issues discussed here do not change. 

We note, however, that we cannot embed our models in the framework
of gauge mediated Supersymmetry breaking (GMSB)
\nref\DiNe{M. Dine and A.E. Nelson,
 Phys. Rev. D48 (1993) 1277, hep-ph/9303230.}%
\nref\DiNS{M. Dine, A.E. Nelson and Y. Shirman,
 Phys. Rev. D51 (1994) 1362, hep-ph/9408384.}%
\nref\DNNS{M. Dine, A.E. Nelson Y. Nir and Y. Shirman,
 Phys. Rev. D53 (1996) 2658, hep-ph/9507378.}%
\nref\GiRa{G.F. Giudice and R. Rattazzi, hep-ph/9801271.}%
\refs{\DiNe-\GiRa}\ with a low breaking scale. 
The reason is that such models predict that the gravitino is lighter than 
the proton. If baryon number is not conserved, the proton decays via
$p\rightarrow G+K^+$. The $\lambda^{\prime\prime}_{112}$ coupling 
contributes to this decay at tree level and is therefore
very strongly constrained 
\nref\CCL{K. Choi, E.J. Chun and J.S. Lee,
 Phys. Rev. D55 (1997) 3924, hep-ph/9611285.}%
\refs{\CCL,\ChKe}:
\eqn\pGK{\lambda^{\prime\prime}_{112}\leq5\times10^{-16}\left(
{\tilde m\over300\ GeV}\right)^2\left({m_{3/2}\over1\ eV}\right).}
All other $\lambda^{\prime\prime}_{ijk}$ couplings 
contribute at the loop level and are constrained as well
\ref\CHL{K. Choi, K. Hwang and J.S. Lee,
 Phys. Lett. B428 (1998) 129, hep-ph/9802323.}.
For $m_{3/2}\sim1\ eV$, the bound \pGK\ would be violated 
(with $\lambda^{\prime\prime}_{112}\sim\lambda^{11}$)
by about eight orders of magnitude.  Therefore, our models of horizontal 
$U(1)$ symmetry broken to lepton parity can be embedded in the GMSB
framework only for $m_{3/2}\gsim10^8\ eV$, that is a Supersymmetry
breaking scale that is higher than $\O(10^8\ GeV)$.

Second, we have not worried about anomaly constraints \IbRo. 
The reason is that these could be satisfied
by extending the matter content of the model and this, again, would
have no effect on the problems of interest to us here.

In the single explicit model that we presented in section 3, our choice
of lepton charges has been dictated by the implications from the 
atmospheric neutrino anomaly and from the MSW solution
of the solar neutrino problem with a small mixing angle.
We emphasize that it is actually simpler to accommodate the
large angle solutions (MSW or vacuum oscillations) of the solar
neutrino problem. We used the small angle option because
we wanted to demonstrate that, first, it can be accommodated in
our framework but that, second, the model does not offer a simplification
in this regard compared to models with integer charges.
The use of half-integer charges in the lepton sector also does not
make significant changes for models using holomorphic zeros to achieve
simultaneously large mixing and large hierarchy \GNS. Finally, we note that
models where some of the $L_i$ and $\bar\ell_i$ carry half-integer
charges and other integer charges do not yield acceptable phenomenology.

\vskip 1 cm
\centerline{\bf Acknowledgements}
We thank Yuval Grossman and Yael Shadmi for useful discussions.
Y.N. is supported in part by the United States -- Israel Binational
Science Foundation (BSF) and by the Minerva Foundation (Munich).
 
\listrefs
\end